\documentstyle[12pt]{article}

\catcode`\@=11
\long\def\@makefntext#1{
\protect\noindent \hbox to 3.2pt {\hskip-.9pt
$^{{\ninerm\@thefnmark}}$\hfil}#1\hfill}		

\def\@makefnmark{\hbox to 0pt{$^{\@thefnmark}$\hss}}  

\def\ps@myheadings{\let\@mkboth\@gobbletwo
\def\@oddhead{\hbox{}
\rightmark\hfil\ninerm\thepage}
\def\@oddfoot{}\def\@evenhead{\ninerm\thepage\hfil
\leftmark\hbox{}}\def\@evenfoot{}
\def\sectionmark##1{}\def\subsectionmark##1{}}

\renewcommand{\thefootnote}{\fnsymbol{footnote}}

\newcounter{sectionc}\newcounter{subsectionc}\newcounter{subsubsectionc}
\renewcommand{\section}[1] {\vspace*{0.6cm}\addtocounter{sectionc}{1}
\setcounter{subsectionc}{0}\setcounter{subsubsectionc}{0}\noindent
	{\normalsize\bf\thesectionc. #1}\par\vspace*{0.4cm}}
\renewcommand{\subsection}[1] {\vspace*{0.6cm}\addtocounter{subsectionc}{1}
	\setcounter{subsubsectionc}{0}\noindent
	{\normalsize\it\thesectionc.\thesubsectionc. #1}\par\vspace*{0.4cm}}
\renewcommand{\subsubsection}[1]
{\vspace*{0.6cm}\addtocounter{subsubsectionc}{1}
	\noindent {\normalsize\rm\thesectionc.\thesubsectionc.
		\thesubsubsectionc.#1}\par\vspace*{0.4cm}}

\newcounter{appendixc}
\newcounter{subappendixc}[appendixc]
\newcounter{subsubappendixc}[subappendixc]

\renewcommand{\appendix}[1] {\vspace*{0.6cm}
        \refstepcounter{appendixc}
        \setcounter{figure}{0}
        \setcounter{table}{0}
        \setcounter{equation}{0}
        \renewcommand{\thefigure}{\Alph{appendixc}.\arabic{figure}}
        \renewcommand{\thetable}{\Alph{appendixc}.\arabic{table}}
        \renewcommand{\theappendixc}{\Alph{appendixc}}
        \renewcommand{\theequation}{\Alph{appendixc}.\arabic{equation}}
        \noindent{\bf Appendix \theappendixc #1}\par\vspace*{0.4cm}}

\def\abstracts#1{{

\centering{\begin{minipage}{12.2truecm}\footnotesize\baselineskip=12pt\noindent
	\centerline{\footnotesize ABSTRACT}\vspace*{0.3cm}
	\parindent=0pt #1
	\end{minipage}}\par}}

\renewenvironment{thebibliography}[1]
	{\begin{list}{\arabic{enumi}.}
	{\usecounter{enumi}\setlength{\parsep}{0pt}
\setlength{\leftmargin 1.25cm}{\rightmargin 0pt}
	 \setlength{\itemsep}{0pt} \settowidth
	{\labelwidth}{#1.}\sloppy}}{\end{list}}

\newcounter{itemlistc}
\newcounter{romanlistc}
\newcounter{alphlistc}
\newcounter{arabiclistc}

\newcommand{\fcaption}[1]{
        \refstepcounter{figure}
        \setbox\@tempboxa = \hbox{\footnotesize Fig.~\thefigure. #1}
        \ifdim \wd\@tempboxa > 6in
           {\begin{center}
        \parbox{6in}{\footnotesize\baselineskip=12pt Fig.~\thefigure. #1}
            \end{center}}
        \else
             {\begin{center}
             {\footnotesize Fig.~\thefigure. #1}
              \end{center}}
        \fi}

\newcommand{\tcaption}[1]{
        \refstepcounter{table}
        \setbox\@tempboxa = \hbox{\footnotesize Table~\thetable. #1}
        \ifdim \wd\@tempboxa > 6in
           {\begin{center}
        \parbox{6in}{\footnotesize\baselineskip=12pt Table~\thetable. #1}
            \end{center}}
        \else
             {\begin{center}
             {\footnotesize Table~\thetable. #1}
              \end{center}}
        \fi}

\def\@citex[#1]#2{\if@filesw\immediate\write\@auxout
	{\string\citation{#2}}\fi
\def\@citea{}\@cite{\@for\@citeb:=#2\do
	{\@citea\def\@citea{,}\@ifundefined
	{b@\@citeb}{{\bf ?}\@warning
	{Citation `\@citeb' on page \thepage \space undefined}}
	{\csname b@\@citeb\endcsname}}}{#1}}

\newif\if@cghi
\def\cite{\@cghitrue\@ifnextchar [{\@tempswatrue
	\@citex}{\@tempswafalse\@citex[]}}
\def\citelow{\@cghifalse\@ifnextchar [{\@tempswatrue
	\@citex}{\@tempswafalse\@citex[]}}
\def\@cite#1#2{{$\null^{#1}$\if@tempswa\typeout
	{IJCGA warning: optional citation argument
	ignored: `#2'} \fi}}

 1
 1
 1

\font\ninerm=cmr9

\begin{document}

\newcommand{\st}{\scriptstyle}
\newcommand{\sst}{\scriptscriptstyle}
\newcommand{\mco}{\multicolumn}
\newcommand{\epp}{\epsilon^{\prime}}
\newcommand{\vep}{\varepsilon}
\newcommand{\ppg}{\pi^+\pi^-\gamma}
\newcommand{\vp}{{\bf p}}
\newcommand{\ko}{K^0}
\newcommand{\kb}{\bar{K^0}}
\newcommand{\al}{\alpha}
\newcommand{\ab}{\bar{\alpha}}
\def\z2{\ifmmode Z_2\else $Z_2$\fi}
\def\ie{{\it i.e.},}
\def\eg{{\it e.g.},}
\def\etc{{\it etc}}
\def\etal{{\it et al.}}
\def\ibid{{\it ibid}.}
\def\gev{\,{\rm GeV}}
\def\to{\rightarrow}
\def\epem{\ifmmode e^+e^-\else $e^+e^-$\fi}
\def\Re{{\cal R \mskip-4mu \lower.1ex \hbox{\it e}\,}}
\def\Im{{\cal I \mskip-5mu \lower.1ex \hbox{\it m}\,}}
\def\be{\begin{equation}}
\def\ee{\end{equation}}
\def\bea{\begin{eqnarray}}
\def\eea{\end{eqnarray}}
\def\CPbar{\hbox{{\rm CP}\hskip-1.80em{/}}}

\rightline{\vbox{\halign{&#\hfil\cr
&hep-ph/9503265\cr}}}
\rightline{\vbox{\halign{&#\hfil\cr
&UCD-95-7\cr}}}

\centerline{{\normalsize\bf Helicity Probabilities For Heavy Quark
Fragmentation}}
\centerline{{\normalsize\bf Into Heavy-light Excited Mesons}
\footnote{Work supported by the Department of
Energy, contract DE-FG03-91ER40674.}
\footnote{Presented at the {\it Beyond the Standard Model -- IV},
Granlibakken -- Tahoe City, CA, December 13-18, 1994.}
}
\baselineskip=15pt

\centerline{\footnotesize Tzu Chiang Yuan}
\baselineskip=13pt
\centerline{\footnotesize\it Davis Institute for High Energy Physics}
\baselineskip=12pt
\centerline{\footnotesize\it Department of Physics, University of California,
CA 95616, USA}
\centerline{\footnotesize Internet address: yuantc@ucdhep.ucdavis.edu}

\vspace*{0.9cm}
\abstracts{
After a brief review on how heavy quark symmetry constraints the
helicity fragmentation probabilities for a heavy quark hadronizes
into heavy-light hadrons, we present a heavy quark fragmentation model
to extract the value for the Falk-Peskin probability $w_{3/2}$
describing the fragmentation of a heavy quark into a heavy-light meson whose
light degrees of freedom have angular momentum ${3 \over 2}$.
We point out that this probability depends on the longitudinal momentum
fraction $z$ of the meson and on its transverse momentum $p_\bot$
relative to the jet axis. In this model, the light degrees
of freedom prefer to have their angular momentum aligned transverse to,
rather than along, the jet axis.
Implications for the production of excited heavy mesons,
like $D^{**}$ and $B^{**}$, are briefly discussed.
}

\normalsize\baselineskip=15pt
\setcounter{footnote}{0}
\renewcommand{\thefootnote}{\alph{footnote}}

\bigskip

Heavy quark $Q$ produced at a high energy reaction can come with a very
high degree of polarization.
For example, at the $Z$ resonance the charm and
bottom quarks are predicted to be 67 and 94\% left-handed polarized,
respectively, in the Standard Model.  An interesting
question to ask is to what extend this large initial polarization of the
heavy quark $Q$ can be retained during the process of fragmentation.
This hadronization process involves redistribution of energy at a
smaller scale and thus nonperturbative QCD effects are overwhelmingly
important. Since heavy quark spin and flavor are decouple from the dynamics
at the limit $M_Q \to \infty$, heavy quark effective theory
provides a powerful tool to analyze this problem. Indeed, Falk and
Peskin \cite{falkpeskin} pointed out recently that heavy quark symmetry
can impose very tight constraints on the helicity probabilities of
heavy-light hadrons produced by the fragmentation/hadronization of a heavy
quark.

Heavy quark symmetry implies heavy-light hadrons can be classified into
$SU(2)$ doublets $(H,H^*)$ with total angular momentum
$(s,s^*) = (j_l - {1 \over 2},  j_l + {1 \over 2})$ respectively,
where $j_l$ is the eigenvalue of the angular momentum $(J_l)$ of the light
degrees of freedom. For instance, in the quark model picture
of a heavy-light meson, one can write $J_l = S_{\bar q} + L$ where
$S_{\bar q}$ is the spin of the antiquark and $L$ is the orbital angular
momentum. Let $P^{H}_{Q,h_Q \to s,h}$ denote the probability
for a heavy quark $Q$ with helicity $h_Q$ (along the fragmentation axis) to
fragment into a heavy-light hadron $H$,
with angular momentum of the light degrees of freedom $j_l$, total
angular momentum $s$, and total helicity $h$.
Let $p_{j_l}(h_l)$ denote the probability
for the heavy quark to fragment into the light degrees of freedom with
angular momentum $j_l$ and helicity $h_l$. Then
$P^{H}_{Q,h_Q \to s,h}$ is given by a incoherent sum of
$p_{j_l}(h_l)$ over all possible $h_l$ weighted by the square of the
Clebsch-Gordan coefficient
$\langle s_Q,h_Q;j_l,h_l \vert s,h \rangle$
\cite{falkpeskin,wisetalk}
\begin{equation}
\label{mastereq}
P^{H}_{Q,h_Q \to s,h} \propto \sum_{h_l} p_{j_l}(h_l)
\vert \langle s_Q,h_Q;j_l,h_l \vert s,h \rangle \vert^2 \; .
\end{equation}
An identical formula holds for $H^*$ as well and will therefore be omitted.
We will call $\{ p_{j_l}(h_l) \}$ the Falk-Peskin probabilities. They
are conditional probabilities  and satisfy the following
constraints:
$(i) \, 0  \leq p_{j_l}(h_l)  \leq 1$, $(ii)\, \sum_{h_l} p_{j_l}(h_l) = 1$,
and $(iii) \, p_{j_l}(h_l) = p_{j_l}(-h_l)$ (Parity invariance). Therefore
the number of independent Falk-Peskin probabilities
is equal $j_l - {1 \over 2}$ for mesons and $j_l$ for baryons.

Now we would like to demonstrate how Eq.(\ref{mastereq}) works by studying
several examples. For simplicity, we will assume the heavy quark
is purely left-handed and denote $P^{H}_{Q,-{1 \over 2} \to s,h}$ and
$P^{H^*}_{Q,-{1 \over 2} \to s^*,h}$ by $P^{H}(h)$ and
$P^{H^*}(h)$ respectively. \\
(1) $\underline{j_l=0}$. In this case, all the heavy quark spin transfers to
the spin ${1 \over 2}$ baryon $\Lambda_Q$. The Falk-Peskin probability
is therefore trivial. \\
(2) $\underline{j_l^P={1 \over 2}^{\pm}}$. This implies
$(s,s^*)^P$ = $(0^-,1^-)$ or $(0^+,1^+)$.
For the odd parity, the doublet $(H,H^*)$ is the usual
$(D,D^*)$ or $(B,B^*)$ multiplet.
For the even parity, these are the $(D_0^*,D_1^\prime)$ and
$(B_0^*,B_1^\prime)$ multiplets that have not been identified
experimentally. The Falk-Peskin probabilities are completely fixed by
parity invariance:
$(p_{1/2}(-{1 \over 2}),p_{1/2}({1 \over 2})) =({1 \over 2},{1 \over 2})$.
Therefore the helicity fragmentation probabilities for the doublet
are given by
\begin{equation}
\label{p01}
\left(
\begin{array}{cc}
P^{H^*}(h) \\  \\ P^{H}(h)
\end{array}
\right)
=
\left(
\begin{array}{ccc}
{1 \over 2} & {1 \over 4} &  0 \\
 & & \\
 & {1 \over 4} &
\end{array}
\right)  \; \; ,
\end{equation}
where the helicity $h$ runs through the values $-1$, 0, +1 across the table.
One consequence of this result is that all the heavy quark spin
information is lost in this case \cite{falkpeskin}. \\
(3) $\underline{j_l = 1}$. In this case, we have
$(s,s^*)=({1 \over 2},{3 \over 2})$ and the
doublet $(H,H^*)$ can be identified as
$(\Sigma_Q,\Sigma_Q^*)$ with $Q \, = \, b$ or $c$ quark. There
is one nontrivial Falk-Peskin probability $w_1$
cannot be determined by heavy quark symmetry:
$\{p_1(h_l)\} = ({1 \over 2}w_1,1-w_1,{1 \over 2}w_1)$
where $h_l$ runs through the values $-1$, 0, 1 from left to right.
The helicity fragmentation probabilities for the
doublet can be worked out easily:
\begin{equation}
\label{p21}
\left(
\begin{array}{cc}
P^{H^*}(h) \\  \\ P^{H}(h)
\end{array}
\right)
=
\left(
\begin{array}{cccc}
{1 \over 2}w_1 & {2 \over 3}(1-w_1) &  {1 \over 6}w_1 & 0 \\
 & & & \\
 & {1 \over 3} (1-w_1) & {1 \over 3}w_1 &
\end{array}
\right)  \; \; ,
\end{equation}
where $h$ runs through the values $-{3 \over 2}$, $-{1 \over 2}$,
$+{1 \over 2}$, $+{3 \over 2}$ across the table.
We note that $w_1$ can be expressed
in terms of the helicity fragmentation probabilities of
the spin ${1 \over 2}$ state as follows:
\begin{equation}
w_1 = { P^{H}({1 \over 2})
\over P^{H}({1 \over 2}) + P^{H}(-{1 \over 2}) } \; .
\end{equation}
(4) $\underline{j_l = {3 \over 2}}$. This implies $(s,s^*)=(1,2)$. The doublet
$(H,H^*)$ can be identified as $D^{**}=(D_1,D^*_2)$ or $B^{**}=(B_1,B^*_2)$.
There is also one nontrivial Falk-Peskin probability $w_{3/2}$
cannot be determined by heavy quark symmetry:
\begin{equation}
\label{p32}
p_{3/2} ( h_l ) =
\left( {1 \over 2}w_{3/2},{1 \over 2}(1 - w_{3/2}),
 {1 \over 2}(1 - w_{3/2}),{1 \over 2}w_{3/2} \right) \; \; ,
\end{equation}
where the helicity $h_l$ of the light degrees of freedom
runs through the values $-{3 \over 2}$, $-{1 \over 2}$,
$+{1 \over 2}$, $+{3 \over 2}$ across the table.
The helicity fragmentation probabilities are given by
\begin{equation}
\label{pdoublet}
\left(
\begin{array}{cc}
P^{H^*}(h) \\  \\ P^{H}(h)
\end{array}
\right)
=
\left(
\begin{array}{ccccc}
{1 \over 2} w_{3/2} & {3 \over 8}(1 - w_{3/2}) &
{1 \over 4} (1 - w_{3/2}) & {1 \over 8} w_{3/2} & 0 \\
 & & & & \\
 & {1 \over 8}(1 - w_{3/2}) & {1 \over 4}(1 - w_{3/2}) &
  {3 \over 8}w_{3/2} &
\end{array}
\right)  \; \; ,
\end{equation}
with the helicity $h$ runs through the values $-2$, $-1$, 0, +1, +2
across the table. $w_{3/2}$ can therefore be expressed
in terms of the helicity fragmentation probabilities of the spin 1
state as follows:
\begin{equation}
w_{3/2} = \frac{P^H(-1) + P^H(1) - {1 \over 2} P^H(0)}
{P^H(-1) + P^H(1) + P^H(0)}  \; .
\end{equation}

In Ref.[3], we presented a heavy quark fragmentation model that allows
us to extract a perturbative result for $w_{3/2}$. The model is based on the
perturbative calculation of the fragmentation functions for
$b$ quark to fragment into P-wave heavy-heavy $(b \bar c)$ mesons \cite{frag}.
Heavy quark fragmentation functions have well-defined limit as
the heavy quark mass goes to infinity \cite{w32}. The Falk-Peskin
probability $w_{3/2}$ can be generalized to $w_{3/2}(y)$,
$w_{3/2}(t)$, and  $w_{3/2}(y,t)$ that can depend on the kinematical variables
$y$ and $t$ of the produced meson. $y$ and $t$ are the scaling variables
corresponding to the longitudinal momentum fraction $z$ and the transverse
momentum $p_\bot$ of the mesons with respect to the fragmentation axis,
respectively, according to the following relations,
\begin{equation}
y = { 1  - (1 - r) z  \over z r} \; , \;\;\; {\rm and} \;\;\;
t = {|{\bf p}_\bot| \over r (m_b+m_c)} \; , \;\;\; {\rm with} \;\;\;
r={m_c \over m_b+m_c} \; .
\end{equation}
In the heavy quark limit $r \to 0$, simple analytic expressions for
$w_{3/2}(y,t)$, $w_{3/2}(t)$, and $w_{3/2}(y)$ were derived \cite{w32}: \\
${\rm            } \;\;\;\;\; \; \; \; w_{3/2} (y,t) =$
\begin{equation}
{3 y^2 (y-1)^2 t^2 [ 4y^2 + y(y+4)t^2 + t^4 ] \over
(t^2 + y^2)  [ y^4(y-4)^2 + y^3(24+y-4y^2+y^3)t^2
+ y^2 (17-2y+2y^2)t^4 + (1+y)^2 t^6 ] } \; ;
\end{equation}
\begin{equation}
\label{w32t}
w_{3/2} (t) = {9 \over 80} \, \cdot \, \left(
{n_1 + n_2 \, {\rm Arctan} \, (t) + n_3 \, {\rm ln} \, (1+t^2)
\over d_1 + d_2 \, {\rm Arctan} \, (t) + d_3 \, {\rm ln} \,
(1+t^2)} \right) \; ,
\end{equation}
with
$n_1 =  t (630 - 2605  t^2 + 231  t^4)$,
$n_2 =  - 5 (126 - 735   t^2 + 160  t^4 + 5  t^6)$,
$n_3 =   - 1280  t (1 - t)(1 + t)$,
$d_1  =  t ( 105 - 812 t^2 + 79 t^4)$,
$d_2 =  - 3 ( 35 - 355 t^2 + 93 t^4 + 3 t^6)$, and
$d_3 = - 144 t(2 - 3 t^2)$; and
\begin{equation}
\label{w32y}
w_{3/2}(y) = {1 \over 10} \, \cdot \,
{(y-1)^2(12+8y+5y^2) \over (8+4y^2+y^4)} \; .
\end{equation}
It has been shown \cite{w32} that $w_{3/2}(t)$ and $w_{3/2}(y)$
are always less than 0.5 for all values of $y$ and $t$.
The original Falk-Peskin probability is given by
$w_{3/2}={29 \over 114}$, a result first derived by
Chen and Wise \cite{chenwise}.

The above results can be applied to the fragmentation processes
$c \to (D_1,D_2^{*})$ and $b \to (B_1,B_2^{*})$.
The prediction of $w_{3/2} \leq 0.5$ in this model implies that
light degrees of freedom with helicities $h_l = \pm {1 \over 2}$ always have
a larger population than the  maximum helicity states of
$h_l = \pm {3 \over 2}$.
This prediction supports the speculation of Falk and Peskin \cite{falkpeskin}
that the angular momentum of the light degrees of freedom
with $j_l = {3 \over 2}$ prefers to align transverse to,
rather than along, the fragmentation axis.
This spin alignment of the heavy quark can be
detected by anisotropy measurements in the decay products from
the hadronic transitions between the two doublets
$(1^+,2^+)$ and  $(0^-,1^-)$ \cite{falkpeskin}.
Our heavy quark fragmentation model predicts that
these anisotropies vary significantly with $y$ and $t$.

An upper bound of $w_{3/2} \leq 0.24$ at the 90\%
confidence level has been deduced from the existing
experimental data for the charm system \cite{falkpeskin}.
Our prediction of $w_{3/2} = {29 \over 114} \approx 0.25$ suggests
that present experiment may be close to observing a nonzero value for
$w_{3/2}$. We note that the Peterson fragmentation model, widely used
in the literature, contains no spin information  and it
is not consistent with heavy quark symmetry.
It is therefore impossible to calculate $w_{3/2}$ in such a model.
On the other hand, string models of fragmentation
tend to give significantly larger values of $w_{3/2}$ \cite{falkpeskin} and
may already have been ruled out.

We conclude that future measurements of the Falk-Peskin probability $w_{3/2}$
for the charm and bottom systems, including the dependence of $w_{3/2}$
on the scaling variables $y$ and $t$,  can provide valuable insights
into the dynamics of heavy quark fragmentation. Finally, it is also
interesting to extend our heavy quark fragmentation model to the case of
baryon to get a prediction for $w_1$. We will leave this challenge
for more ambitious readers.

\footnotesize
\baselineskip=10pt

\bibliographystyle{unsrt}

\begin{thebibliography}{99}
\bibitem{wisetalk}
{This compact formula for $P^H_{Q,h_Q \to s,h}$ is due to Mark Wise.}
\end{thebibliography}

\end{document}